\documentclass[prd,twocolumn]{revtex4}
\usepackage{graphicx, epsfig}
\usepackage{color}

\newcommand{\be}{\begin{equation}}
\newcommand{\ee}{\end{equation}}
\newcommand{\bea}{\begin{eqnarray}}
\newcommand{\eea}{\end{eqnarray}}

\newcommand{\gapp}{\mathrel{\raise.3ex\hbox{$>$}\mkern-14mu
              \lower0.6ex\hbox{$\sim$}}}
\newcommand{\lapp}{\mathrel{\raise.3ex\hbox{$<$}\mkern-14mu
              \lower0.6ex\hbox{$\sim$}}}

\begin{document}
\title{Non-local (but also non-singular) physics at the last stages of gravitational collapse}
\author{Anshul Saini, Dejan Stojkovic}
\affiliation{ HEPCOS, Department of Physics, SUNY at Buffalo, Buffalo, NY 14260-1500, USA}


\begin{abstract}
We study the end stages of gravitational collapse of the thin shell of matter in ingoing Eddington-Finkelstein coordinates. We use the functional Schrodinger formalism to capture quantum effects in the near singularity limit. We find that that the equations of motion which govern the behavior of the collapsing shell near the classical singularity become strongly non-local. This reinforces previous arguments that quantum gravity in the strong field regime might be non-local. We managed to solve the non-local equation of motion for the dust shell case, and found an explicit form of the wavefunction describing the collapsing shell. This wavefunction and the corresponding probability density are non-singular at the origin, thus indicating that quantization should be able to rid gravity of singularities, just as it was the case with the singular Coulomb potential.
\end{abstract}


\maketitle

\section{Introduction}
What happens at the last stages of the gravitational collapse of some distribution of matter is still unknown. The reason is our lack of a fully fledged theory of quantum gravity  which will fatefully describe quantum dynamics in very strong gravitational fields (e.g. near classical singularities). Since the formulation of quantum gravity still seems to be far from our reach, we have to work with what we have at hand, and try to push it as far as possible. Along the way, we might get a glimpse of what the ultimate theory of quantum gravity should look like.

The purpose of this paper is to study quantum aspects of gravitational collapse of a shell of matter in the context of the functional Schrodinger formalism \cite{Vachaspati:2006ki,Vachaspati:2007hr,Benedict:1996qy,Kiefer:1990pt,Vachaspati:2007ur,Greenwood:2008zg,Greenwood:2010mr,Greenwood:2008ht,Wang:2009ay,Halstead:2011gs,
Greenwood:2010uy,Greenwood:2009pd,Greenwood:2009gp,Greenwood:2010sx}. We will work in  Eddington-Finkelstein coordinates which are convenient for studying the question of the black hole formation till the very end where the collapsing matter distribution crosses its own Schwarzschild radius and starts approaching the classical singularity at the center. The first interesting finding is that the equations of motion describing behavior of the collapsing shell near the classical singularity become non-local. It has been argued for a while that (for various reasons) quantum gravity should ultimately be a manifestly non-local theory \cite{Greenwood:2008ht,Wang:2009ay,Horowitz:2003he,Giddings:2006sj,Giddings:2006be}, (see also \cite{Hehl:2009es,Chicone:2012cc,Mashhoon:2013jqa}). Our finding is a strong indication that something like that might indeed be true. While the functional Schrodinger formalism is not a full theory of quantum gravity, it should however capture some aspects of it. Non-locality might be one of those important aspects.

Non-local equations are notoriously difficult to solve. However, manipulating the equations of motion in the near singularity limit, we managed to find an explicit solution to the non-local Schrodinger equation. Interestingly enough, the solution for the wavefunction is non-singular at the origin. In fact, the probability density becomes zero exactly at the origin. This indicates that quantization can perhaps remove classical singularities from gravity, as argued from many different points of view \cite{Greenwood:2008ht,Wang:2009ay,Bogojevic:1998ma,Guendelman:2009vz,Rasouli:2013sda,Olmo:2012nx,Vakili:2011nc,Casadio:2013uga,McInnes:2008fk}.

In section \ref{metric} we setup the metric for the collapsing dust shell in ingoing Eddington-Finkelstein coordinates and derive transformations that we will use later. In section \ref{hamiltonian} we apply Gauss-Codazzi method to find the conserved quantity which has a clear interpretation of the Hamiltonian. In section \ref{dsq} we quantize this Hamiltonian using the functional Schrodinger formalism in the near singularity limit. As we mentioned, the functional Schrodinger equation is non-local. In the same section we solve this equation explicitly to find the wavefunction and probability density (which is non-singular). In section \ref{dw} we repeat the same procedure for the shell whose energy density is constant (i.e. spherically symmetric domain wall). In section \ref{dwq} we quantize the domain wall shell and find again that the equation governing the behavior near the origin is non-local.
We do not explicitly solve for the wavefuction in this case since the expressions are cumbersome. Finally, we give conclusions in section \ref{conclusions}.

\section{The metric in ingoing Eddington-Finkelstein coordinates}
\label{metric}
In this section we will setup the metric of a collapsing shell of matter in ingoing Eddington-Finkelstein coordinates. Since this space-time foliation is  non-singular at the Schwarzschild radius, it will allow us to study the the gravitational collapse as the shell is approaching  the classical singularity at the center.

The radius of a collapsing spherically symmetric shell of mater is $R$. The parameter of evolution is the ingoing null coordinate $v$ related to the asymptotic Schwarzschild time  as
\be
  v=t+r^*
\ee
where $r^*$ is the tortoise coordinate. The trajectory of the collapsing shell is then simply $r=R(v)$.
The metric outside the collapsing shell is
\be
  ds^2=-\left(1-\frac{R_s}{r}\right)dv^2+2dvdr+r^2d\Omega^2, \hspace{2mm} r>R(v).
  \label{out_Met}
\ee
By Birkhoff theorem, the interior metric is Minkowski
\be
ds^2 = - dT^2 + dr^2 + r^2 {\Omega}^2 ,   r< R(v)
\ee
The interior time coordinate, $T$, is related to the ingoing null coordinate, $v$, via the proper time on the shell, $\tau$. [Note that the proper time $\tau$ is different from the same quantity in Schwarzschild coordinates since the space-time foliation is different.]  The relations are
\be
  \frac{dT}{d\tau}=\sqrt{1+\left(\frac{dR}{d\tau}\right)^2}
  \label{dTdtau}
\ee
and
\be
  \frac{dv}{d\tau}=\frac{1}{B}\left(\frac{dR}{d\tau}-\sqrt{B+\left(\frac{dR}{d\tau}\right)^2}\right)
  \label{dvdtau}
\ee
where
\be
  B\equiv 1-\frac{R_s}{R}.
\ee
 From Eq.~(5), we can get,
\be
{R_{\tau}}^2=\frac{B}{{(1-\frac{B}{R_v})}^2-1}
\ee
From Eq.~(4) and Eq.~(5),
\be \label{vT}
\frac{dv}{dT}=\frac{1}{\sqrt{{R_v}^2-2R_v+B}}
\ee
We will use these relations when needed to convert from one to another time coordinate.

\section{Collapse of the dust shell: conserved quantity}
\label{hamiltonian}

In this section we will derive a quantity which remains conserved during the collapse of the dust (pressureless) shell. We will then identify this quantity with the Hamiltonian. We will use the Gauss-Codazzi method, also known as Israel approach for surface layers (see e.g. \cite{Ipser:1983db,himanshu,schmidt} and references therein).

Let us consider a shell which is a time-like three-surface separating two regions of space-time. The extrinsic curvature, $K_{ij}$, in this case is discontinuous. The explicit form of the extrinsic curvature can be derived from Einstein equations as
\be
[{K^i}_j] = 8\pi \left({S^i}_j - \frac{1}{2} {{\partial}^i}_j {S^k}_k \right)
\ee
 where $S$ is surface stress-energy defined as
\be
{S^{\alpha}}_{\beta} = \lim_{\epsilon \to 0} \int^{+\epsilon}_{-\epsilon} {T^{\alpha}}_{\beta} \,dn
\ee
One can show that, in general,
\be
{{S_i}^m}_{\mid m} + [ {T^n}_i] = 0
\ee
For our case of the thin shell of dust the surface stress energy can be written as
\be
S^{\alpha \beta} = \sigma u^{\alpha} u^{\beta} ,
\ee
where $\sigma$ is energy per unit area of the shell. For the constant rest mass of the shell, $\sigma$ is not constant, and changes its value as the radius of the shell changes.
Our extrinsic curvature tensor becomes
\be
[K_{ij}] = 8 \pi \sigma \left( u_i u_j + \frac{1}{2} {}^{(3)} g_{ij}\right) .
\ee
Outside the shell the stress energy tensor $[{T^n}_i]$ will be zero. Then Eq.~(11) gives
\be
\frac{d\sigma}{d \tau}  + \sigma {u^m}_{\mid m} = 0 .
\ee

The metric on the shell is
\be
ds^2 = - d{\tau}^2 + R^2 (\tau) ( d{\theta}^2 + {\sin}^2 \theta d {\phi}^2)
\ee
where $\tau$ is the proper time parameter of the observer located on the shell.
For this form of the metric  Eq.~(14) becomes
\be
\frac{{(\sigma {({}^{(3)}g)}^{\frac{1}{2}}u^i)}_{,i}}{{({}^{(3)}g) }^{\frac{1}{2}}} = 0
\ee
 Since $u^{\tau}=1$ and  ${}^{(3)}g = R^4 (\tau)$ it becomes $4 \pi R^2 \sigma = \mu$. We can think of $\mu$ as the rest mass of the shell which always remains constant for the dust shell.  Now we can use Eq.~(13) to find equations of motion
\be
[K_{\theta \theta}] = 8 \pi \sigma \left( u_{\theta} u_{\theta} + \frac{1}{2} {}^{(3)}g_{\theta \theta} \right) = \mu
\ee
One can find $K_{\theta \theta}$ as
\be
K_{\theta \theta} = - n_{\theta;\theta} = - \frac{1}{2} n^2 g_{\theta \theta , r} = - r n^r
\ee
Combining Eq.~(17) and Eq.~(18) we get
\be
- r (n^{r+} - n^{r-}) = \mu
\ee
where $n^{r+}, n^{r-}$ are radial components toward exterior and interior region respectively.  The components of $u$ and $n$ can be evaluated by using the dot products $n.u=0$, $n.n=1$ and $u.u=-1$. Towards the exterior region of shell, we get
\be
n_v u^v +n_r u^r = 0
\ee
\be
\left(1-\frac{R_s}{r}\right) {(u^v)}^2 - 2 u^v u^r = 1
\ee
\be
\left(1-\frac{R_s}{r}\right) {(n_r)}^2 - 2 n_r n_v = 1
\ee
We can solve this system of equations to get $n_r= \pm u^v$. Substituting this result in Eq.~(21) gives
\be
\left(1-\frac{R_s}{r}\right) {(n_r)}^2 - 2  u^r n_r = 1 .
\ee
Now, $n^r = g^{r \nu} n_{\nu}$, which leads to
\be
n^r =   \left(1-\frac{R_s}{r}\right) n_r +n_v .
\ee
Substituting $n_v$ from Eq.~(22), we get
\be
n^r = \frac{1+\left(1-\frac{R_s}{r}\right) {n_r}^2}{2 n_r} .
\ee
Using Eq.~(23) and Eq.~(25), we find
\be
n^r = \pm \sqrt{\left(1-\frac{R_s}{r}\right) + {(u^r)}^2} .
\ee
We will ignore negative sign because at the shell $u^r = \dot{R}$, so the relation becomes
\be
n^{r+} = \sqrt{\left(1-\frac{R_s}{R}\right) + {\dot{R}}^2}  .
\ee
We can easily obtain $n^{r-}$ from $n^{r-}$ by substituting $M=0$
\be
n^{r-} = \sqrt{1 + {\dot{R}}^2} .
\ee
Substituting $n^{r+}$ and $n^{r-}$ into Eq.~(19), we get
\be
-r \left( \sqrt{\left(1-\frac{R_s}{R}\right) + {\dot{R}}^2} -  \sqrt{1 + {\dot{R}}^2}\right) = \mu
\ee
which can be simplified to give
\be \label{M}
M=\mu{{\left(1+{R_\tau}^2\right)}^{\frac{1}{2}}} - \frac{{\mu}^2}{2R}
\ee
The quantity $M$ in Eq.~(\ref{M}) is an integral of motion, and has a clear interpretation of the total energy. It contains the rest mass of the shell $\mu$, the kinetic energy represented by $R_\tau$, and gravitational self-energy $\mu^2/(2R)$. We will therefore identify it with the Hamiltonian of the system.

We now express the $R_\tau$ in terms of $R_T$ to obtain
\be \label{MT}
M=\mu{\left(\frac{1}{\sqrt{1-{R_T}^{2}}} - \frac{\mu}{2R}\right)}
\ee
We can then write down an effective action that gives Eq.~(\ref{MT}) as its Hamiltonian as
\be
S_{eff}= -\mu\int\left(\sqrt{1-{R_T}^2}-\frac{\mu G}{2R}\right)
\ee
The ultimate goal is to find an action in terms of the ingoing coordinate $v$, so we write
\be
S_{eff}= -\mu\int\left(\sqrt{1-{\left({{\frac{dR}{dv}\frac{dv}{dT}}}\right)}^2}-\frac{\mu G}{2R}\right) .
\ee
Substituting  $dv/dT$ from Eq.(8), we arrive at the desired action
\be
S_{eff}=-\int dv \mu\left(\sqrt{B-2R_v}-\frac{\mu G}{2R}\sqrt{{R_v}^2 - 2 R_v+B}\right) .
\ee
The corresponding Lagrangian is
\be \label{Lv}
L = - \mu\left(\sqrt{B-2R_v}-\frac{\mu G}{2R}\sqrt{{R_v}^2 - 2 R_v+B}\right) .
\ee
We needed an explicit form of the Lagrangian so that we can define the canonical momentum and find the Hamiltonian in terms of momentum.
Canonical momentum  in this case is defined as $\Pi=\frac{\partial L}{\partial R_v}$, which yields
\be \label{momentum}
\Pi = \mu\left(\frac{1}{\sqrt{B-2R_v}} + \frac{\mu G}{2R}\frac{(R_v - 1)}{\sqrt{{R_v}^2 - 2 R_v+B}}\right) .
\ee
Finally,  Hamiltonian corresponding to the Lagrangian in Eq.~(\ref{Lv})
\be \label{H}
H = \mu(B-R_v)\left(\frac{1}{\sqrt{B-2 R_v}} - \frac{\mu G}{2R}\frac{1}{\sqrt{{R_v}^2 - 2 R_v+B}}\right) .
\ee
where the velocity $R_v$ should be eliminated by the use of Eq.~(\ref{momentum}). We emphasis that so far we did not use any approximations, so the Hamiltonian in Eq.~(\ref{H}) is  exact.

\section{Quantum collapse of the dust shell in the limit  of $R\rightarrow 0$}
\label{dsq}
The main goal of the this paper is to see what happens at the last stages of the collapse of the shell, i.e. when $R\rightarrow 0$. Since we have an explicit Hamiltonian of the system, we can apply the functional Schrodinger formalism and study quantum effects near the classical singularity. In the framework of the functional Schrodinger formalism, we will simply write down the Schrodinger equation for the wave-functional $\Psi[R(v)]$, and try to solve it.

We first derive the behavior of $R_\tau$ near $R=0$. From Eq.~(\ref{M}), we have
\be
R_{\tau} = \sqrt{{\left(\frac{M}{\mu}+\frac{\mu G}{2R}\right)}^2 -1} .
\ee
From here we see that  $R_{\tau}\approx\frac{\mu G}{2R}$ as $R$ is approaching zero. Substituting this result in Eq.(7), we find
\be \label{Rv}
R_v\approx-\frac{1}{2}{\left(\frac{\mu G}{R}\right)}^2
\ee
Thus, the rate at which the dust shell collapses near $R=0$ diverges as $R_v \propto \frac{1}{R^2}$.

In this limit the Hamiltonian in Eq.~(\ref{H}) can be approximated as
\be \label{H0}
H = \mu(-R_v)\left[\frac{1}{\sqrt{2 \mid R_v \mid}} - \frac{\mu G}{2R}\frac{1}{R_v}\right]
\ee
which gives
\be
R_v=2{\left(\frac{H}{\mu}-\frac{G\mu}{2R}\right)}^2
\ee
For $R\rightarrow0$, we can ignore the constant term $H/\mu$, and we will again get  Eq.~(\ref{Rv}).

In the limit $R\rightarrow0$, the canonical momentum reduces to
\be
\Pi=\mu\left[\frac{1}{\sqrt{2\mid R_v \mid}}+\frac{\mu  G}{2R}\right]
\ee
Expressing $R_v$ in terms of $\Pi$ in Hamiltonian (\ref{H0}) we get
\be \label{H0P}
H=\frac{-R}{G}{\left[1-\frac{2\Pi R}{{\mu}^2 G}\right]}^{-1} +\frac{{\mu}^2 G}{2R} .
\ee
The Hamiltonian in Eq.~(\ref{H0P}) governs the evolution of the collapsing dust shell in vicinity of $R=0$.
As in the standard quantization procedure, we promote the momentum $\Pi$ into an operator
\be
\Pi = -i \hbar \frac{\partial }{\partial R} .
\ee
We can now write the functional Schrodinger equation for the wave-functional $\psi [R(v)]$
\be \label{se}
H \psi = i \hbar \frac{\partial \psi }{\partial v}
\ee
and try to solve it. Unfortunately, the structure of the Hamiltonian (\ref{H0P}) is such that the usual treatment is practically impossible.
The main problem is that the differential operator in Hamiltonian (\ref{H0P}) is non-local. This finding represents a strong support for suggestions that quantum gravity
might be ultimately a non-local theory.

While finding solutions to non-local equations is very difficult, we will show that it is possible to define a procedure (similar to the one outlined in \cite{Vachaspati:2007ur}) which will lead to the solution of Eq.~(\ref{se}).
We first isolate the non-local operator $\hat{A}$ from the Hamiltonian (\ref{H0P})
\be
\hat{A}={\left[1-\frac{2\Pi R}{{\mu}^2 G}\right]}^{-1}
\ee
Its inverse is
\be
{\hat{A}}^{-1} = 1- \frac{2\Pi R}{{\mu}^2 G}
\ee
We can take care of the operator ordering as
\be
\hat{A}={\left[1-\frac{1}{{\mu}^2 G}(\hat{\Pi} R+R\hat{\Pi})\right]}^{-1} ,
\ee
so that
\be
{\hat{A}}^{-1}=1-\frac{1}{{\mu}^2 G}(\hat{\Pi} R+R\hat{\Pi}) .
\ee
In terms of derivatives, ${\hat{A}}^{-1}$ is
\be
{\hat{A}}^{-1}=(1+\frac{i}{{\mu}^2 G})+\frac{2 i R}{{\mu}^2 G}\frac{\partial}{\partial R}
\ee
Let's define the action of an operator $\hat{A}$ as $\varphi={\hat{A}}\psi$, which means $\psi={\hat{A}}^{-1}\varphi$, where $\varphi$ is just some function which gives the wavefunction $\psi$ upon action of the operator $\hat{A}$.  Explicit action of ${\hat{A}}^{-1}$ on $\varphi$ converts the equation ${\hat{A}}^{-1}\varphi = \psi$ into a linear differential equation
\be
\frac{d\varphi}{dR}+\frac{1}{2R}(1-i {\mu}^2 G)\varphi+\frac{i  {\mu}^2 G}{2R}\psi =0
\ee
This equation can be solved to give
\be
\varphi = - \frac{i{\mu}^2 G}{2} \frac{\int{R^{-\frac{(1+i  {\mu}^2 G)}{2}}}\psi dR}{R^{\frac{(1- i  {\mu}^{2} G)}{2}}}
\ee
Since $\varphi= \hat{A}\psi$ we obtain the action of $\hat{A}$ as
\be
\hat{A} = - \frac{i{\mu}^2 G}{2} \frac{\int{R^{-\frac{(1+i  {\mu}^2 G)}{2}}}(.)dR}{R^{\frac{(1- i  {\mu}^{2} G)}{2}}}
\ee
where $(.)$ is the placeholder for the function on which $\hat{A}$ is acting.


Let's concentrate on the stationary solutions to Eq.~(\ref{se}) in the form of
\be
\psi (R,v) = \psi (R) e^{iEv/{\hbar}}
\ee
where $v$ is the time evolution parameter, and $E$ is the energy eignevalue.
The time independent Schrodinger equation becomes $H \psi  = E \psi$. The Hamiltonian in Eq.~(\ref{H0P}) in terms of the operator $\hat{A}$ becomes
\be
H  = \frac{R \hat{A}}{G} + \frac{{\mu}^{2} G}{2R}
\ee
Accounting for the ordering of operators,  this Hamiltonian becomes
\be
H  = -\frac{1}{2G} \left(R \hat{A} + \hat{A}R\right)+ \frac{{\mu}^{2} G}{2R} .
\ee
The Schrodinger equation (\ref{se}) becomes
\be
\frac{-1}{2G}\left (R \hat{A} + \hat{A}R\right)\psi+ \frac{{\mu}^{2} G}{2R}\psi = E\psi
\ee
 When $\hat{A}$ operates on $R$ we get
\be
\hat{A} (R) = -{\frac{\alpha}{2}  R^{-\frac{1}{2}(1- \alpha)}\int{R^{-\frac{1}{2}(1+ \alpha)} R dR}}
\ee
which yields
\be
\hat{A}(R) = \frac{-\alpha R + \beta}{3- \alpha}
\ee
where $\alpha = i {\mu}^2 G$ and $\beta$ is an integration constant. So our equation becomes
\be
\frac{-1}{2G}\left (R \hat{A} \psi + \frac{\alpha R - \beta}{ \alpha- 3} \psi \right)+ \frac{{\mu}^{2} G}{2R}\psi = E\psi
\ee
Now we can move all the terms to one side and separate the term with the integral
\be
\int{R^{-\frac{1}{2}(1+ \alpha)} \psi dR} = \frac{4G R^{-\frac{1+ \alpha}{2}}}{\alpha} \left[ \frac{1}{2G}\left(\frac{\alpha R - \beta}{\alpha - 3}\right) -\frac{{\mu}^2 G }{2R} + E\right] \psi
\ee
We can now differentiate this equation with respect to $R$ to remove integration. Differentiation yields
\begin {widetext}

\bea
&&\left[\frac{2R^{\frac{1}{2}(1- \alpha)}}{\alpha -3}  - 2 {\mu}^2 G^2 R^{-\frac{1}{2}(3+ \alpha)}+\left(\frac{4GE}{\alpha} - \frac{2 \beta}{\alpha( \alpha - 3)}\right) R^{-\frac{1}{2}(1+ \alpha)}\right] \psi'  =  \\
&& \left[ \left(\frac{\alpha-1}{\alpha - 3}  +1\right) R^{-\frac{1}{2}(1+ \alpha)} - {\mu}^2 G^2 (\alpha+3) R^{-\frac{1}{2}(5+ \alpha)} + \left( \frac{2 G E (\alpha + 1)}{\alpha} - \frac{ \beta (\alpha +1 )}{\alpha (\alpha - 3)}\right)R^{-\frac{1}{2}( 3 + \alpha)} \right] \psi \nonumber
\eea

\end{widetext}
This can be written as
\be
\frac{d \psi}{\psi} = \int{\frac{a_1 + a_2 R +a_3 R^2}{a_4 R + a_5 R^2 + a_6 R^3}dR}
\ee
where
$a_1 =- {\mu}^{2} G^2 (\alpha+3)$ , $a_2  = \left(\frac{2GE (\alpha+1)}{\alpha} - \frac{\beta (\alpha+1) }{\alpha (\alpha - 3)}\right)$, $a_3  = 1+ \frac{\alpha - 1}{\alpha + 3}$, $a_4 = -2{\mu}^2 G^2$, $a_5 =\left(\frac{4GE}{\alpha} - \frac{2 \beta}{\alpha (\alpha - 3)}\right) $ and $a_6 = \frac{2}{\alpha - 3}$
This integral can be solved for general values of constants. However, since we are working in the limit of  $R \approx 0$, we keep only the leading order terms
\be
\ln \psi  =  \int{\frac{a_1}{a_4 R } dR} + {\rm constant}
\ee
Solving this equation and substituting the values of the constants, we find the solution for the wavefunction
\be
\psi =  \lambda R^{\frac{3+i {\mu}^2 G}{2}} .
\ee
where $\lambda $ is a constant. The corresponding probability density $P = {\psi}^{*}\psi$ is
\be \label{pd}
{\mid \psi \mid}^2 = {\lambda}^2 R^3 .
\ee
This result is very important. It demonstrates that the probability density associated with the wavefunction $\psi$ which describes the collapse of the shell of matter is non-singular near the classical singularity. In fact, the probability density in Eq.~(\ref{pd}) vanishes exactly at $R=0$. It is remarkable that a simple quantum treatment of the gravitational collapse indicates that classical singularity at the center can be removed.

\section{Collapse of an infinitely thin spherical domain wall: conserved quantity}
\label{dw}
It is a logical possibility that the non-local behavior in the near singularity region that we found in the previous section is an artifact of the example that we were studying, i.e. the dust shell of matter. In this section we will repeat the procedure for the shell of matter whose energy per unit area, $\sigma$, is constant, which is the situation represented by a spherically symmetric domain wall. We will find that the near singularity behavior is qualitatively the same even in this case, i.e. the Hamiltonian becomes non-local in $R \rightarrow 0$ limit.

In the case of the domain wall, mass $M$ is also a conserved quantity. However, the relation relating mass with $R$ and $R_\tau$ is now given as
\be \label{conserved}
M = \frac{1}{2}\left({ \sqrt{{1+{R_{\tau}}^2}}+ \sqrt{{B+{R_{\tau}}^2}}}\right) 4 \pi \sigma R^2
\ee
where $B \equiv 1-2GM/R$.
This expression is implicit as it contains $M$ in $B$. The explicit relation of $M$ in terms of $R_\tau$ and $R$ can be written as
\be
M = 4 \pi \sigma R^2 \left[{\sqrt{1- {R_\tau}^2}}- 2 \pi G \sigma R\right]
\ee
Using the relation between $T$ and $\tau$ (Eq.~(4)), we get
\be
M = 4 \pi \sigma R^2 \left[\frac{1}{\sqrt{1- {R_T}^2}}- 2 \pi G \sigma R\right]
\ee
 The effective action which can reproduce above relation, i.e. gives the correct mass conservation law, can be written as
\be
S_{eff}= -4 \pi \sigma \int R^{2} \left[\sqrt{1-{R_T}^2}- 2\pi G \sigma R \right]
\ee
We want to convert the $T$ coordinate into the infalling $v$ coordinate as
\be
S_{eff}= -4 \pi \sigma \int R^{2} \left[\sqrt{1-{\left({\frac{\partial{R}}{\partial{v}}\frac{\partial{v} }{\partial{T}}}\right)}^2}- 2\pi G \sigma R\right]
\ee
After substituting the expression for $\frac{dv}{dT}$, we arrive at
\be
S_{eff} = -4 \pi \sigma \int dv R^{2} \left(\sqrt{B-2R_{v}}-2 \pi \sigma R \sqrt{{R_v}^{2}-2 R_{v}+B}\right) .
\ee
The corresponding Lagrangian can be written as
\be
L_{eff}=  -4 \pi \sigma R^{2} \left[\sqrt{B-2R_{v}}-2 \pi \sigma R \sqrt{{R_v}^{2}-2 R_{v}+B}\right]
\ee
Canonical momentum is defined as $\Pi = \frac{\partial L}{\partial R_{v}}$, so we obtain
\be
\Pi = -4 \pi \sigma R^{2} \left[ \frac{-1}{\sqrt{B-2R_v}}-\frac{2 \pi \sigma GR(R_v -1 )}{\sqrt{{R_v}^{2}-2 R_{v}+B}}\right]
\ee
Since the Hamiltonian is $ H =(\pi R_v - L )$,  substituting the $R_v$ in terms of $\Pi$ gives
\be\label{HDW}
H= 4 \pi \sigma R^{2} (B-R_{v})\left[\frac{1}{\sqrt{B-2R_v}}- \frac{2 \pi \sigma G R}{\sqrt{{R_v}^{2}-2 R_{v}+B}}\right]
\ee

\section{Quantum collapse of the domain wall in the limit of $R\rightarrow 0$}
\label{dwq}

Since we have an explicit Hamiltonian of the system, we can apply the functional Schrodinger formalism again and study quantum effects near the classical singularity.
Again from the conserved quantity $M$ we can derive
\be
\mid R_{\tau} \mid = \sqrt{{\left( \frac{M}{4 \pi \sigma R^{2}}+ 2 \pi \sigma G R\right)}^{2} - 1}
\ee
From here we see that  $R_v \approx - \frac{2 M^2}{{(4 \pi \sigma R^2)}^2}$  as $R$ is approaching zero.
Thus, the rate at which the dust shell collapses near $R=0$ diverges as $R_v \propto \frac{1}{R^4}$, in contrast with the dust shell where the divergence was quadratic.

In the same limit the Hamiltonian in Eq.~(\ref{HDW}) can be approximated as
\be
H =  4 \pi \sigma R^2 (-R_v)\left[ \frac{1}{\sqrt{2\mid R_v \mid}} + \frac{2 \pi \sigma G R}{\mid R_v \mid} \right]
\ee
 Solving this equation for $R_v$ yields  $\mid R_v \mid \approx \frac{2 h^2}{R^4}$ where $h= \frac{H}{4 \pi \sigma}$ which is the same as the above derived behavior.

The exact form of canonical momentum can be written as
\be
\Pi = -4 \pi \sigma R^{2} \left[ \frac{-1}{\sqrt{B-2R_v}}-\frac{2 \pi \sigma GR(R_v -1 )}{\sqrt{{R_v}^{2}-2 R_{v}+B}} \right]
\ee
which in $R\rightarrow 0$ limit can be approximated as
\be
\Pi = 4 \pi \sigma R^{2} \left[ \frac{1}{\sqrt{2\mid R_v \mid}}+ 2 \pi \sigma GR \right]
\ee
From this we can get $R_v$ in terms of $\pi$ as
\be \label{rv}
R_v = - \frac{1}{2} {\left[\frac{\pi}{4 \pi \sigma R^{2}}-2\pi \sigma GR\right]}^{-2}
\ee
The Hamiltonian given by Eq.(\ref{HDW}) can now be approximated as
\be
H = 4\pi \sigma R^{2} (-R_v) \left[\frac{1}{\sqrt{B-2 R_v}}\right]
\ee
Submitting the value of $R_v$ from Eq.~(\ref{rv}) gives
\be\label{HDW0P}
H = - \frac{R}{\left(1-\frac{\pi}{8{\pi}^3 {\sigma}^2 G R^{3}}\right)}
\ee
The Hamiltonian in Eq.~(\ref{HDW0P}) governs the evolution of the collapsing spherical domain wall in vicinity of $R=0$.
As before, we promote the momentum $\Pi$ into an operator
\be
\Pi = -i \hbar \frac{\partial }{\partial R} .
\ee
We can now write the functional Schrodinger equation
\be \label{sedw}
H \psi = i \hbar \frac{\partial \psi }{\partial v}
\ee
and try to solve it. However, as in our previous case of the dust shell, we see that the differential operator in Hamiltonian (\ref{HDW0P}) is non-local.
This fact reinforces an indication that quantum gravity should be ultimately a non-local theory.

We will now follow the procedure we outlined in the section (\ref{dsq}).
We isolate the non-local operator
\be
\hat{A}={\left[1-\frac{\Pi}{8{\pi}^3 {\sigma}^2 G R^{3}}\right]}^{-1}
\ee
For convenience, we set the constant $\alpha = \frac{1}{16 {\pi}^2 {\sigma}^2 G}$ which gives
\be
\hat{A} =\frac{1}{ \left(1 - 2 \alpha \frac{\Pi}{R^3}\right) }
\ee
Now, we can define
\be
{\hat{A}}^{-1} = 1 - 2\alpha \frac{\Pi}{R^3}
\ee
We take care of the ordering problem as
\be
{\hat{A}}^{-1} = 1- \alpha \left[\Pi \frac{1}{R^3} + \frac{1}{R^3} \Pi\right]
\ee
which also makes this operator unitary.
In terms of derivatives, we have
\be
{\hat{A}}^{-1} = 1+ i \alpha \left(\frac{2}{R^3}\frac{\partial}{\partial R} - \frac{3}{R^4}\right)
\ee
Let us again define $\psi = {\hat{A}}^{-1} \varphi$, which leads to differential equation
\be
\frac{\partial \varphi}{\partial R} + \left(\frac{R^3}{2i \alpha}- \frac{3}{2R}\right) \varphi - \frac{R^3}{2i \alpha} \psi =0
\ee
This linear differential equation can be solved to give
\be
 \varphi = \frac{R^{3/2} e^{-\frac{R^4}{8i\alpha}} }{2i \alpha} \int R^{3/2} e^{\frac{R^4}{8i \alpha}} \psi dR
\ee
Since $\varphi = \hat{A}\psi$, this gives the operator $\hat{A}$ as
\be
 \hat{A} = \frac{R^{3/2} e^{-\frac{R^4}{8i\alpha}} }{2i \alpha} \int R^{3/2} e^{\frac{R^4}{8i \alpha}} (.) dR
\ee
where $(.)$ is the placeholder for the function on which $\hat{A}$ is acting.
In principle,  one can follow the procedure we outlined  in section \ref{dsq} and solve the non-local Schrodinger equation like. However, in this case calculations are  much more cumbersome because of presence of additional exponential in the operator $\hat{A}$ and will not be shown here.

\section{Conclusions}
\label{conclusions}
In this paper we studied quantum aspects of the gravitational collapse near the classical singularity as seen by an infalling observer. Since gravity is the by far the weakest force in nature, we expect that quantum mechanics will significantly modify classical behavior of gravity only in the strong field regimes, e.g. near classical singularities. In the absence of a fully fledged theory of quantum gravity, we worked in the context of the functional Schrodinger formalism applied to a simple gravitational system - collapsing shell of matter.
We used the Eddington-Finkelstein space-time foliation which is convenient for studying the question of the black hole formation till the very end where the collapsing shell crosses its own Schwarzschild radius and starts approaching the classical singularity at the center.
We derived the conserved quantity with the clear interpretation as the Hamiltonian of the system and quantized the theory. In the $R \rightarrow 0$  limit, we found that the equation which describes the quantum evolution of the collapsing shell is strongly non-local. Non-local terms which are usually suppressed in large distance limit, become dominant in the near singularity limit.  This conforms some earlier speculations and related studies. As an important step forward, we managed to solve this non-local equation explicitly and found the form of the wavefuction. Remarkably,  the wavefunction and its corresponding probability density are non-singular at  $R \rightarrow 0$. This is an indication that quantization can remove classical singularities from gravity, just as it was the case with the singular electromagnetic Coulomb potential.

\vskip.2cm {\bf Acknowledgment} \vskip.2cm
 DS acknowledges the financial support from NSF, award number PHY-1066278.



\begin{thebibliography}{99}




\bibitem{Vachaspati:2006ki}
  T.~Vachaspati, D.~Stojkovic and L.~M.~Krauss,
  Phys.\ Rev.\  D {\bf 76}, 024005 (2007)
  [arXiv:gr-qc/0609024];

\bibitem{Vachaspati:2007hr}
  T.~Vachaspati and D.~Stojkovic,
  Phys.\ Lett.\  B {\bf 663}, 107 (2008)
  [arXiv:gr-qc/0701096].

\bibitem{Benedict:1996qy}
  E.~Benedict, R.~Jackiw and H.~J.~Lee,
  Phys.\ Rev.\ D {\bf 54}, 6213 (1996)
  [hep-th/9607062].

\bibitem{Kiefer:1990pt}
  C.~Kiefer and T.~P.~Singh,
  Phys.\ Rev.\ D {\bf 44}, 1067 (1991).


\bibitem{Vachaspati:2007ur}
  T.~Vachaspati,
  Class.\ Quant.\ Grav.\  {\bf 26}, 215007 (2009)
  [arXiv:0711.0006 [gr-qc]].




\bibitem{Greenwood:2008zg}
  E.~Greenwood and D.~Stojkovic,
  JHEP {\bf 0909}, 058 (2009)
  [arXiv:0806.0628 [gr-qc]].



\bibitem{Greenwood:2010mr}
  E.~Greenwood, D.~C.~Dai and D.~Stojkovic,
  Phys.\ Lett.\ B {\bf 692}, 226 (2010)
  [arXiv:1008.0869 [astro-ph.CO]].


\bibitem{Greenwood:2008ht}
  E.~Greenwood and D.~Stojkovic,
  JHEP {\bf 0806}, 042 (2008)
  [arXiv:0802.4087 [gr-qc]].


\bibitem{Wang:2009ay}
  J.~E.~Wang, E.~Greenwood and D.~Stojkovic,
  Phys.\ Rev.\ D {\bf 80}, 124027 (2009)
  [arXiv:0906.3250 [hep-th]].

\bibitem{Halstead:2011gs}
  E.~Halstead,
  JCAP {\bf 1308}, 043 (2013)
  [arXiv:1106.2279 [gr-qc]].



\bibitem{Greenwood:2010uy}
  E.~Greenwood,
  arXiv:1002.2433 [gr-qc].


\bibitem{Greenwood:2009pd}
  E.~Greenwood,
  JCAP {\bf 1001}, 002 (2010)
  [arXiv:0910.0024 [gr-qc]].


\bibitem{Greenwood:2009gp}
  E.~Greenwood, E.~Halstead and P.~Hao,
  JHEP {\bf 1002}, 044 (2010)
  [arXiv:0912.1860 [gr-qc]].

\bibitem{Greenwood:2010sx}
  E.~Greenwood, D.~I.~Podolsky and G.~D.~Starkman,
  JCAP {\bf 1111}, 024 (2011)
  [arXiv:1011.2219 [gr-qc]].








\bibitem{Horowitz:2003he}
  G.~T.~Horowitz and J.~M.~Maldacena,
  JHEP {\bf 0402}, 008 (2004)
  [arXiv:hep-th/0310281].




\bibitem{Giddings:2006sj}
  S.~B.~Giddings,
  Phys.\ Rev.\  D {\bf 74}, 106005 (2006)
  [arXiv:hep-th/0605196].


\bibitem{Giddings:2006be}
  S.~B.~Giddings,
  Phys.\ Rev.\  D {\bf 74}, 106009 (2006)
  [arXiv:hep-th/0606146].


\bibitem{Hehl:2009es}
  F.~W.~Hehl and B.~Mashhoon,
  Phys.\ Rev.\ D {\bf 79}, 064028 (2009)
  [arXiv:0902.0560 [gr-qc]].

\bibitem{Chicone:2012cc}
  C.~Chicone and B.~Mashhoon,
  Phys.\ Rev.\ D {\bf 87}, no. 6, 064015 (2013)
  [arXiv:1210.3860 [gr-qc]].

\bibitem{Mashhoon:2013jqa}
  B.~Mashhoon,
  Class.\ Quant.\ Grav.\  {\bf 30}, 155008 (2013)
  [arXiv:1304.1769 [gr-qc]].


\bibitem{Bogojevic:1998ma}
  A.~Bogojevic and D.~Stojkovic,
  Phys.\ Rev.\  D {\bf 61}, 084011 (2000)
  [arXiv:gr-qc/9804070];


\bibitem{Guendelman:2009vz}
  E.~Guendelman, A.~Kaganovich, E.~Nissimov and S.~Pacheva,
  Int.\ J.\ Mod.\ Phys.\ A {\bf 25}, 1571 (2010)
  [arXiv:0908.4195 [hep-th]].





\bibitem{Rasouli:2013sda}
  S.~M.~M.~Rasouli, A.~H.~Ziaie, J.~Marto and P.~V.~Moniz,
  arXiv:1309.6622 [gr-qc].


\bibitem{Olmo:2012nx}
  G.~J.~Olmo and D.~Rubiera-Garcia,
  Phys.\ Rev.\ D {\bf 86}, 044014 (2012)
  [arXiv:1207.6004 [gr-qc]].


\bibitem{Vakili:2011nc}
  B.~Vakili,
  Int.\ J.\ Theor.\ Phys.\  {\bf 51}, 133 (2012)
  [arXiv:1102.1682 [gr-qc]].

\bibitem{Casadio:2013uga}
  R.~Casadio, O.~Micu and F.~Scardigli,
  arXiv:1311.5698 [hep-th].

\bibitem{McInnes:2008fk}
  B.~McInnes,
  Nucl.\ Phys.\ B {\bf 807}, 33 (2009)
  [arXiv:0806.3818 [hep-th]].



\bibitem{Ipser:1983db}
  J.~Ipser and P.~Sikivie,
  Phys.\ Rev.\  D {\bf 30}, 712 (1984).


\bibitem{schmidt}
H.\ -J.\  Schmidt,
Gen.\ Relat.\  Grav.\ {\bf 16},  1053 (1984)
[arXiv:gr-qc/0105106].

\bibitem{himanshu}
H.\  Kumar, S.\ Alam, S.~Ahmad,
Gen.\  Relat.\  Grav.\  {\bf 45},  125 (2013)
[arXiv:1211.0128].



\end{thebibliography}
\end{document}